\documentclass[journal, hidelinks]{IEEEtran}
\usepackage[switch]{lineno}
\usepackage{cite}
\usepackage{amsmath,amssymb,amsfonts,mathtools}
\usepackage{algorithmic}
\usepackage{graphicx}
\usepackage{textcomp}
\usepackage{xcolor}
\usepackage{orcidlink}
\def\BibTeX{{\rm B\kern-.05em{\sc i\kern-.025em b}\kern-.08em
    T\kern-.1667em\lower.7ex\hbox{E}\kern-.125emX}}
\usepackage[final]{microtype}
\usepackage[toc, acronym]{glossaries}
\glsdisablehyper
\usepackage{comment}
\usepackage{siunitx}
\usepackage{graphicx}

\usepackage{fancyhdr}

\newcommand{\herm}[1]{\ensuremath{#1^{\mathrm{H}}}}
\newcommand{\trans}[1]{\ensuremath{#1^{\mathrm{T}}}}
\newcommand{\id}[1]{\ensuremath{\mathbf{I}_{#1}}}
\newcommand{\power}[1]{\ensuremath{\mathit{P}_{#1}}}
\newcommand{\euler}{\ensuremath{\mathit{e}}}
\newcommand{\snr}{\ensuremath{\mathrm{SNR}}}
\newcommand{\bsf}[1]{\ensuremath{\boldsymbol{\mathsf{#1}}}}
\newcommand{\complex}{\ensuremath{\mathbb{C}}}
\newcommand{\covmat}[1]{\ensuremath{\mathbf{C}_{#1}}}
\newcommand{\frob}{\ensuremath{\mathrm{F}}}
\DeclareMathOperator{\expec}{\mathrm{E}}
\DeclareMathOperator*{\argmin}{arg\,min}
\DeclareMathOperator*{\argsort}{arg\,sort}
\DeclareMathOperator{\trace}{\mathrm{tr}}
\DeclareMathOperator{\prob}{\mathrm{Pr}}

\newcommand{\ie}{\textit{i.e.} }

\IEEEaftertitletext{\vspace{-1.5\baselineskip}}

\newcommand{\changefont}{\color{blue}\fontsize{9}{9}\selectfont}

\let\oldmaketitle\maketitle
\renewcommand{\maketitle}{%
	\oldmaketitle
	\thispagestyle{fancy}
}

\begin{document}
\newacronym{snr}{SNR}{signal-to-noise ratio}
\newacronym{simo}{SIMO}{single-input multiple-output}
\newacronym{ml}{ML}{maximum likelihood}
\newacronym{llr}{LLR}{log-likelihood ratio}
\newacronym{csi}{CSI}{channel state information}
\newacronym{pam}{PAM}{pulse-amplitude modulation}
\newacronym{im}{IM}{index modulation}
\newacronym{ofdm}{OFDM}{orthogonal frequency-division multiplexing}
\newacronym{ser}{SER}{symbol error rate}
\newacronym{ed}{ED}{energy detector}
\newacronym{hsnr}{HSNR}{high \acrshort{snr}}
\newacronym{abque}{ABQUE}{assisted best quadratic unbiased estimator}
\newacronym{bque}{BQUE}{best quadratic unbiased estimator}
\newacronym{ask}{ASK}{amplitude-shift keying}
\newacronym{awgn}{AWGN}{additive white Gaussian noise}
\newacronym{pim}{PIM}{permutational index modulation}
\newacronym{se}{SE}{spectral efficiency}
\newacronym{pm}{PM}{permutation modulation}
\newacronym{itu}{ITU}{International Telecommunication Union}
\newacronym{embb}{eMBB}{enhanced mobile broadband}
\newacronym{mmtc}{mMTC}{massive machine type communications}
\newacronym{urllc}{URLLC}{ultra-reliable low-latency communications}

    \bibliographystyle{IEEEtran-normspace}
	\bstctlcite{IEEEexample:BSTcontrol}

    \title{\vspace*{-0.3em}Low-Complexity Detection of Permutational Index Modulation for Noncoherent Communications}
    \author{Marc\! Vilà-Insa\,\orcidlink{0000-0002-7032-1411},~\IEEEmembership{Graduate~Student~Member,~IEEE},
    Aniol Martí\,\orcidlink{0000-0002-5600-8541},~\IEEEmembership{Graduate~Student~Member,~IEEE},
    Meritxell~Lamarca\,\orcidlink{0000-0002-8067-6435},~\IEEEmembership{Member,~IEEE},
	and Jaume Riba\,\orcidlink{0000-0002-5515-8169},~\IEEEmembership{Senior~Member,~IEEE}
	\thanks{This work was funded by project MAYTE (PID2022-136512OB-C21) by MICIU/AEI/10.13039/501100011033 and ERDF/EU, grants 2021 SGR 01033 and 2022 FI SDUR 00164 by Departament de Recerca i Universitats de la Generalitat de Catalunya, and grant 2023 FI ``Joan Or\'o'' 00050 by Departament de Recerca i Universitats de la Generalitat de Catalunya and the\nobreak\ ESF+.}%
    \thanks{The authors are with the Departament de Teoria del Senyal i Comunicacions, Universitat Politècnica de Catalunya (UPC), 08034 Barcelona (e-mail: \{marc.vila.insa, aniol.marti, meritxell.lamarca, jaume.riba\}@upc.edu).}}

    \maketitle
    
    \begin{abstract}
        This work presents a massive \acrshort{simo} scheme for wireless communications with one-shot noncoherent detection.
        It is based on \acrlong{pim} over \acrshort{ofdm}.
        Its core principle is to convey information on the ordering in which a fixed collection of values is mapped onto a set of \acrshort{ofdm} subcarriers. 
        A spherical code is obtained which provides improved robustness against channel impairments.
        A simple detector based on the sorting of quadratic metrics of data is proposed.
        By exploiting statistical \acrlong{csi} and hardening, it reaches near-\acrshort{ml} error performance with a low-complexity implementation.
    \end{abstract}
    
        
    \section{Introduction}
        \IEEEPARstart{N}{ext} generation networks are envisioned to accommodate a large number and variety of devices such as mobile phones, sensor networks or even vehicles.
        This heterogeneity means that future systems must be versatile and allow for different use cases, as described by the \acrfull{itu}~\cite{shafi_5g_2017}.
        Furthermore, \acrshort{ofdm} has been selected as the principal waveform of 5G, hence the different technologies developed should be compatible with it~\cite{memisoglu_fading_2019}.

        A prominent operation mode of next generation systems is \acrfull{mmtc}, in which the volume of information to transmit is small but must be highly reliable and fulfill severe latency constraints.
        This paper proposes a low-latency scheme aimed at obtaining reliable communications with low complexity transceivers.
        In particular, it focuses on one-shot detection (\ie extremely low latency) with a noncoherent receiver, which does not require instantaneous \acrfull{csi} and allows to reduce the training overhead entailed by its acquisition.
        To achieve high reliability over the wireless medium, a massive \acrshort{simo} system is considered to exploit the channel hardening property of large arrays~\cite{Jing2016}.

        On the one hand, the proposed scheme can be regarded as a generalization of \acrshort{ofdm} with \acrlong{im} (\acrshort{ofdm}-\acrshort{im})~\cite{Basar2013}.
        It has attracted a great deal of attention in recent years and a lot of research has been focused on improving its \acrfull{se} by employing multi-level constellations~\cite{Wen2021}.
        Most of these contributions have been made in the context of coherent communications, with notable exceptions~\cite{Choi2018,Fazeli2022}.

        On the other hand, the proposed scheme has the structure of a \textit{Variant-I type} of permutation modulation~\cite{Slepian1965}, originally envisioned for the Gaussian channel.
        The codes employed are constructed by permuting the components of a given initial vector, hence conveying information in their order.
        While some works in the literature extend their use to noncoherent schemes with fading~\cite{Choi2018,Fazeli2022}, to the best of the authors' knowledge, none of them accounts for antenna correlation, and only~\cite{Fazeli2022} deals with a multilevel modulation.
        In~\cite{VilaInsa2024}, a general quadratic framework for noncoherent energy detection in massive \acrshort{simo} systems was presented.
        Leveraging these results, we propose a scheme for low-complexity detection of multi-level \acrshort{ofdm}-\acrshort{im} codes based on the ordering of quadratic statistics of data.
        It exploits statistical \acrshort{csi} to reliably detect the transmitted codewords, even under correlated fading.
        Most notably, the \textit{\acrfull{abque}}~\cite{VilaInsa2024} tailored to \acrshort{ofdm}-\acrshort{im} displays close to \acrfull{ml} performance in a variety of communication scenarios with remarkable reduction in computational complexity. 

   	\section{Signal model}
	    Consider a massive \acrshort{simo} point-to-point system with $N$ antennas at the receiver which uses $K$ \acrshort{ofdm} subcarriers.
	    Communication is one-shot and through a fading channel that remains constant for a single \acrshort{ofdm} block.
        At every band $k$, fading is modeled as $\bsf{h}_k\sim\mathcal{CN}(\mathbf{0}_N,\covmat{\bsf{h},k})$, which is a common assumption in both well-established and state-of-the-art channel models~\cite[Sec.~2.2]{bjornson_massive_2017}.
        It is assumed to be independent among different frequency bands,\footnote{This is usually achieved with techniques such as frequency interleaving~\cite{MullerWeinfurtner2002}.} \ie $\expec[\bsf{h}_k\herm{\bsf{h}}_{k'}]=\mathbf{0}_{N\times N}$, $\forall k\neq k'$.
	    The transmitter sends a codeword $\mathbf{x}\triangleq\trans{[x_1,\dots,x_K]}\in\complex^K$ spanning the full $K$ frequency bands: $x_k$ is sent through $\bsf{h}_k$.
	    	
	    The signal at the receiver in band $k$ is modeled as\footnote{Note that the same model would be obtained for a system with $K$ independent temporal channel uses.
        All the derivations considered herein can be applied in such scenario straightforwardly.}
	    \begin{equation}
        	\bsf{y}_k = \bsf{h}_k\mathsf{x}_k + \bsf{z}_k,\quad k=1,\dots,K,\label{eq:sig_mod}
	    \end{equation}
	    being $\bsf{z}_k\sim\mathcal{CN}(\mathbf{0}_N,\power{\bsf{z}}\id{N})$ \acrfull{awgn}.
        We consider a receiver with \textit{statistical \acrshort{csi}}: it is aware of the distributions of $\bsf{h}_k$ and $\bsf{z}_k$ but not of their realizations.

        In a noncoherent setting, instantaneous \acrshort{csi} is unknown at the receiver.
        Treating the channel as a random variable, the likelihood function of the received signal at the $k$th band for a given $x_k$ is~\cite{VilaInsa2024}
	    \begin{equation}
	        \textstyle f_{\bsf{y}\vert x,k}(\mathbf{y}_k)=\exp\bigl(-\herm{\mathbf{y}_k}\covmat{\bsf{y}\vert x, k}^{-1}\mathbf{y}_k\bigr)\bigm/\bigl(\pi^N\lvert\covmat{\bsf{y}\vert x, k}\rvert\bigr),\label{eq:likelihood}
	    \end{equation}
	    where $\covmat{\bsf{y}\vert x, k}\triangleq \lvert x_k\rvert^2\covmat{\bsf{h},k}+\power{\bsf{z}}\id{N}$.
	    Generalizing it to a multiband setting with independent frequency bands results in $f_{\bsf{Y}\vert\mathbf{x}}(\mathbf{Y})=\prod_{k=1}^{K}f_{\bsf{y}\vert x,k}(\mathbf{y}_k)$, where $\mathbf{Y}\triangleq[\mathbf{y}_1,\dots,\mathbf{y}_K]$.

        Noncoherent schemes of the type studied herein are not able to exploit phase information of individual components of $\mathbf{x}$, since the received signal only conveys information on $\{\lvert x_k\rvert^2\}$, as implied in~\eqref{eq:likelihood}.
        Therefore, $\{x_k\}$ is selected from any $L$-ary unipolar \acrfull{pam}, and each component of $\mathbf{x}$ can take an amplitude level from
        \begin{equation}
            \mathcal{A}_L\triangleq\{\sqrt{\varepsilon_1}\triangleq0<\sqrt{\varepsilon_2}<\dots<\sqrt{\varepsilon_L}\},\quad L\geq2.
        \end{equation}

        The error performance of the one-shot noncoherent receiver considered in this paper is insufficient when no further coding is employed~\cite{VilaInsa2024}.
        To provide robustness to channel impairments and achieve a certain coding gain, while maintaining the desired low-latency and low-complexity implementation, we propose the use of a \acrfull{pm}.
            
        \subsection{Permutational index modulation (PIM)}\label{ssec:pim}
            In \acrshort{pm}~\cite{Slepian1965}, each codeword $\mathbf{x}$ in an alphabet $\mathcal{X}$ is constructed by permuting the elements of a reference vector $\underline{\mathbf{x}}\triangleq\trans{[\underline{x}_1,\dots,\underline{x}_K]}$.
            The set of permutations of $\underline{\mathbf{x}}$ (\ie our codebook) is denoted as $\mathcal{X}\triangleq\Sigma_K(\underline{\mathbf{x}})$.
            The successive components of $\underline{\mathbf{x}}$ are the amplitudes $\sqrt{\varepsilon_l}$ from $\mathcal{A}_L$ in ascending order, each repeated $K_l$ times:
            \begin{equation}
                \underline{\mathbf{x}}\triangleq\trans{[\underbracket[1pt]{0,\dots,0}_{K_1},\underbracket[1pt]{\sqrt{\varepsilon_2},\dots,\sqrt{\varepsilon_2}}_{K_2},\dots,\underbracket[1pt]{\sqrt{\varepsilon_L},\dots,\sqrt{\varepsilon_L}}_{K_L}]}.
            \end{equation}
            Note that $\sum_{l=1}^LK_l=K$.
            From the point of view of \acrshort{im}, $K_1$ corresponds to the portion of inactive subcarriers in the \acrshort{ofdm} symbol~\cite[Sec.~1.2.3]{Wen2021}.

            This scheme yields a codebook with cardinality~\cite{Slepian1965}
            \begin{equation}
                \textstyle \rvert\mathcal{X}\lvert = K! \bigm/ \prod_{l=1}^{L}K_l!. \label{eq:codewords}
            \end{equation}
            The redundancy and error correcting capabilities introduced by \acrshort{pim} (for finite $K$ and $L$) are evidenced by comparing $\lvert\mathcal{X}\rvert$ with $L^K$, which is the total amount of possible unstructured codewords.
        	From \eqref{eq:codewords}, the \acrshort{se} of the alphabet can be computed in bits per channel use [bpcu]:
        	\begin{equation}
        		\textstyle\mathrm{R}_{K}(\{K_l\}) \triangleq \frac{1}{K} \log\lvert\mathcal{X}\rvert \!=\! \frac{1}{K} \bigl( \log(K!) -\! \sum_{l=1}^L \log(K_l!) \bigr), \label{eq:rate}
        	\end{equation}
            where $\log(\boldsymbol{\cdot})$ denotes the binary logarithm.
            The code rate is defined as $\mathrm{R}_{K}(\{K_l\})/\log L$.

            It was proved in~\cite{biglieri1975optimum} that $\lim_{K\to\infty} \mathrm{R}_{K}(\{K_l\}) = \mathrm{H}(\{p_l\})$ for \acrshort{pm}, where $\mathrm{H}(\boldsymbol{\cdot})$ is the Shannon entropy and $p_l \triangleq K_l/K$, for $l=1,\dots,L$.
            It then follows that $\mathrm{R}_{K}(\{K_l\})$ is asymptotically maximized under a uniform policy of $p_l = 1/L$.
            An alternative proof for the asymptotic performance is given in the Appendix, in which it is shown that \acrshort{se} approaches $\mathrm{H}(\{p_l\})$ from below.
            Additionally, it is also proved that uniform policy is optimal (in terms of \acrshort{se}) in the non-asymptotic case as well.
            This yields an asymptotic code rate of $\mathrm{H}(\{p_l\})/\log L \leq 1$, which means the uniform policy does not provide any redundancy for large $K$.
            This is further explored in Sec.~\ref{sec:numerical_results}, particularly in Fig.~\ref{fig:rate}.

    \section{Low complexity detection}
        \subsection{General ML detection}
            Given an equiprobable alphabet, it is known that the probability of incorrectly detecting a transmitted codeword is minimized by the \acrshort{ml} detector, \ie the one that maximizes the likelihood function over all possible transmitted codewords.
	        In our setting, it results in
            \begin{equation}
                \widehat{\mathbf{x}}_{\mathrm{ML}} = \argmin_{\mathbf{x}\in\mathcal{X}} \textstyle\sum\limits_{k=1}^{K}\herm{\mathbf{y}_k}\covmat{\bsf{y}\vert x, k}^{-1}\mathbf{y}_k + \ln\lvert\covmat{\bsf{y}\vert x, k}\rvert. \label{eq:ml}
            \end{equation}
            In general, detecting a single codeword involves $O(\lvert\mathcal{X}\rvert K N^2)$ operations to compute the statistics for every hypothesis $\mathbf{x}$ and $O(\lvert\mathcal{X}\rvert)$ additional ones to find the minimum one.
            This complexity may be reduced by applying a layer of pre-processing to the received signals $\{\bsf{y}_k\}$~\cite{VilaInsa2024}.
            Let $\covmat{\bsf{h},k}\triangleq\mathbf{U}_k\mathbf{\Lambda}_k\herm{\mathbf{U}_k}$ be the eigendecomposition of the channel covariance matrix associated with band $k$.
            By working with $\{\bsf{r}_k\triangleq\herm{\mathbf{U}_k}\bsf{y}_k$\}, which are distributed as $(\bsf{r}_k\vert x_k)\sim\mathcal{CN}(\mathbf{0}_N,\lvert x_k\rvert^2\mathbf{\Lambda}_k+\power{\bsf{z}}\id{N})$, the \acrshort{ml} detection problem reduces to
            \begin{equation}
                \label{eq:ml_simple}
                \widehat{\mathbf{x}}_{\mathrm{ML}}\!=\argmin_{\mathbf{x}\in\mathcal{X}} \!\!\textstyle\sum\limits_{\substack{ k=1..K \\ n=1..N}} \frac{\lvert [\mathbf{r}_{k}]_n\rvert^2}{\lvert x_k\rvert^2\lambda_{n,k}\!+\!\power{\bsf{z}}} + \ln\bigl(\lvert x_k\rvert^2\lambda_{n,k}+\power{\bsf{z}}\bigr),
            \end{equation}
            where $\lambda_{n,k}\triangleq[\mathbf{\Lambda}_k]_{n,n}$.
            This expression generalizes the one used in~\cite{VilaInsa2024} and~\cite{Han2022}.
            Detection with~\eqref{eq:ml_simple} is equivalent to~\eqref{eq:ml}, since there is a one-to-one correspondence between $\bsf{r}_k$ and $\bsf{y}_k$.
    
            Applying this pre-processing has an added complexity of $O(KN^2)$, but now computing the statistics requires $O(\lvert\mathcal{X}\rvert KN)$ operations.\footnote{Diagonalizing $\{\covmat{\bsf{h},k}\}$ does not incur significant computational load since fading statistics remain stable over many channel uses~\cite[Sec.~2.2]{bjornson_massive_2017}.}
            Clearly, the complexity of directly evaluating~\eqref{eq:ml_simple} scales proportionally to $\lvert\mathcal{X}\rvert$.
            Nonetheless, this task can be simplified further by applying the \textit{Viterbi algorithm} in a procedure similar to the one described in~\cite{Viterbo2003}.
            Unfortunately, the total number of states in the trellis representation is $\prod_{l=1}^L(K_l+1)$, which greatly penalizes schemes constructed over large \acrshort{pam} constellations.
            Therefore, true \acrshort{ml} detection of \acrshort{pim} might become infeasible for large $K$ and $L$.
            
            Some special cases in which \acrshort{ml} detection admits low-complexity implementations are now considered.

    	\subsection{Isotropic channel}\label{ssec:iso}
            Under the simplest channel correlation model (\ie the isotropic channel), each frequency band is distributed as $\bsf{h}_k\sim\mathcal{CN}(\mathbf{0}_N,\id{N})$.
            In this case, the resulting \acrshort{ml} detector simplifies to
    		\begin{equation}
    			\widehat{\mathbf{x}}_{\mathrm{ML}} = \argmin_{\sigma\in\Sigma_K} \textstyle\sum\limits_{k=1}^{K} \frac{\lVert\mathbf{r}_k\rVert^2}{\lvert \underline{x}_{\sigma(k)}\rvert^2+\power{\bsf{z}}} + N\ln(\lvert \underline{x}_{\sigma(k)}\rvert^2+\power{\bsf{z}}),\label{eq:ml_iso}
    		\end{equation}
            where $\sigma$ is an index permutation.
    		The logarithmic term is common for all codewords in the alphabet, since they are permutations of the same elements.
    		Therefore, it can be removed from the minimization problem, leaving
    		\begin{equation}
    			\widehat{\mathbf{x}}_{\mathrm{ML}} = \argmin_{\sigma\in\Sigma_K} \textstyle\sum\limits_{k=1}^{K} u_{k}v_{\sigma(k)} \equiv \displaystyle\argmin_{\sigma\in\Sigma_K} \textstyle\sum\limits_{k=1}^{K} u_{\sigma(k)}v_{k},
    		\end{equation}
    		where $u_i\triangleq\lVert\mathbf{r}_i\rVert^2\equiv\lVert\mathbf{y}_i\rVert^2$ and $v_i\triangleq(\lvert \underline{x}_i\rvert^2+\power{\bsf{z}})^{-1}$.
    		This means that \acrshort{ml} detection simplifies to finding the permutation $\sigma\in\Sigma_K$ that minimizes the sum $\sum_{k=1}^{K}u_{\sigma(k)}v_{k}$.
    		
    		This problem can be solved via the \textit{rearrangement inequality}~\cite[Proposition~6.A.3]{Marshall2011}.
            Consider $\{v_k\}$ are arranged in decreasing order, \ie $v_1> v_2>\dots>v_K$.
            Then, the minimum sum is obtained with the permutation that yields the terms $\{u_k\}$ in increasing order.
            This implies \acrshort{ml} detection is equivalent to pairing the most energetic symbols with the most energetic received signal per band.
            More formally, it is expressed as
            \begin{equation}
                \varsigma = \textstyle\argsort_{\sigma\in\Sigma_K}\{\lVert\mathbf{y}_k\rVert^2\},\label{eq:argsort}
            \end{equation}
            where $\argsort$ is a function that sorts the values of $\{\lVert\mathbf{y}_k\rVert^2\}$ and returns $\varsigma $, the permutation that achieves such order.
            The \acrshort{ml} decision is then obtained as
            \begin{equation}
                \arraycolsep=1.2pt\def\arraystretch{1.2}
                \begin{array}{rcccccccl}
                     \bigl\{ & \lVert\mathbf{y}_{\varsigma (1)}\rVert^2 &<& \lVert\mathbf{y}_{\varsigma (2)}\rVert^2 &<& \dots &<& \lVert\mathbf{y}_{\varsigma (K)}\rVert^2 & \bigr\}\\
                     & \Updownarrow && \Updownarrow && && \Updownarrow & \\
                    \trans{\widehat{\mathbf{x}}_{\mathrm{ML}}} = \bigl[ & \underline{x}_{\varsigma ^{-1}(1)} &,& \underline{x}_{\varsigma ^{-1}(2)} &,& \dots &,& \underline{x}_{\varsigma ^{-1}(K)} & \bigr].
                \end{array}\label{eq:simple_ml_sort}
            \end{equation}
            An analogous outcome was observed in~\cite{Slepian1965} for the coherent detection of \acrshort{pm} over the \acrshort{awgn} channel.
            The results in this section (namely,~\eqref{eq:argsort} and~\eqref{eq:simple_ml_sort}) extend it to the noncoherent setting under isotropic fading.
            Using standard sorting algorithms, this procedure involves $O(KN)+O(K\ln K)$ operations, which is a significant reduction compared to the complexity of the general \acrshort{ml} implementation.

            It is worth noting that \acrshort{ml} detection with~\eqref{eq:argsort},~\eqref{eq:simple_ml_sort} yields an additional benefit: it can assess decision reliability at almost no extra cost.
            Such information can be quantified by comparing the likelihood of the most likely decision, $\widehat{\mathbf{x}}_{\mathrm{ML}}$, and the second most probable one, $\widehat{\mathbf{x}}_{2}$, whose permutation is $\varsigma_{2}$:
            \begin{equation}
                \textstyle\mathrm{LLR}=\ln\frac{\prob(\bsf{x}=\widehat{\mathbf{x}}_{\mathrm{ML}}\vert\mathbf{Y})}{\prob(\bsf{x}=\widehat{\mathbf{x}}_{\mathrm{2}}\vert\mathbf{Y})} = \sum_{k=1}^{K} (u_{\varsigma_{2}(k)} - u_{\varsigma (k)}) v_k. \label{eq:llr}
            \end{equation}
            These two codewords only differ in a single permutation of two symbols of consecutive power.
            This allows a very simple computation of~\eqref{eq:llr}:
            \begin{equation}
                \textstyle\mathrm{LLR}= \min_{l\in\mathcal{L}} (u_{\varsigma (l+1)}-u_{\varsigma (l)})(v_{l} - v_{l+1}),
            \end{equation}
            where $\mathcal{L}\triangleq\{\sum_{i=1}^{j}K_i,\,j=1,\dots,L-1\}$.

        \subsection{High SNR regime under general antenna correlation}\label{ssec:general}
            The previous low-complexity \acrshort{ml} detector is only applicable under isotropic fading, which is an unlikely scenario when using large arrays~\cite{Bjoernson2016}.
            In the general case~\eqref{eq:ml_simple}, \acrshort{ml} detection does not admit a simplified implementation based on sorting.
            A notable exception arises when the \acrshort{snr} grows without bounds, defining $\snr\triangleq\bigl(\sum_k\expec[\lvert\mathsf{x}_k\rvert^2]\expec[\lVert\bsf{h}_k\rVert^2]\bigr)\big/\bigl(\sum_k\expec[\lVert\bsf{z}_k\rVert^2]\bigr)$.
            Assuming the receiver is aware of the indices of inactive subcarriers (\ie the ones with transmit power $\varepsilon_1=0$), which form set $\mathcal{K}_1$, we can study the limit of~\eqref{eq:ml_simple} when noise power vanishes:
            \begin{equation}
                \smashoperator{\lim_{\power{\bsf{z}}\to0}}\widehat{\mathbf{x}}_{\mathrm{ML}} = \argmin_{\mathbf{x}\in\mathcal{X}} \textstyle\sum\limits_{k\notin\mathcal{K}_1} \tfrac{1}{\lvert x_k\rvert^2}\herm{\mathbf{r}_k}\mathbf{\Lambda}_k^{-1}\mathbf{r}_k + \ln\bigl\lvert\lvert x_k\rvert^2\mathbf{\Lambda}_k\bigr\rvert.
            \end{equation}
            Once again, the logarithmic terms are removed from the minimization since they are common to all codewords.
            We are left with a problem similar to the one encountered in Sec.~\ref{ssec:iso}.
            Following the same rationale, \acrshort{ml} detection now consists in pairing the most energetic symbols with the highest values of metrics $\{\herm{\mathbf{r}_k}\mathbf{\Lambda}_k^{-1}\mathbf{r}_k\}$, which are quadratic statistics of data.

            This simplification accounts for channel correlation but is only optimal in the high \acrshort{snr} regime and assuming inactive subcarriers have been reliably detected.
            Nevertheless, such a detector can be used suboptimally under any \acrshort{snr} and to detect every subcarrier, regardless of whether they are active or not.
            This sets the basis for the family of detectors introduced next.
            
        \subsection{Proposed detection scheme}
            The previous analyses provide valuable insights about the \acrshort{ml} detection problem~\eqref{eq:ml_simple}.
            They motivate the derivation of a low-complexity detector based on the ordering of some relevant metrics $\{\widehat{\beta}_k(\bsf{r}_k)\}$ obtained from the received signal at each subcarrier $k$:
            \begin{equation}
                \widetilde{\varsigma}= \textstyle\argsort_{\sigma\in\Sigma_K}\{\widehat{\beta}_k(\bsf{r}_k)\}.
            \end{equation}
            More precisely, we propose to use the framework developed in~\cite{VilaInsa2024}, in which the authors devise a family of energy estimators based on quadratic statistics of data.
            While suboptimal in general, when properly designed, these methods can effectively leverage statistical \acrshort{csi} to provide accurate estimations of the transmitted energy in each subcarrier.
            Particularized to our setting, they take the form $\widehat{\beta}_k(\bsf{r}_k)\triangleq\herm{\bsf{r}_k}\mathbf{A}_k\bsf{r}_k-\trace(\mathbf{A}_k)$, which is the estimated energy level from subcarrier $k$ and $\{\mathbf{A}_k\}$ are diagonal matrices that satisfy $\trace(\mathbf{A}_k\mathbf{\Lambda}_k)=1$, so that estimation is \textit{conditionally unbiased}~\cite{VilaInsa2024}.

            As expected, the choice of weighting matrices $\{\mathbf{A}_k\}$ greatly impacts detection performance.
            Within the context of this work, we consider three designs from~\cite{VilaInsa2024}:
            \begin{enumerate}
                \item \textit{Energy detector (\acrshort{ed}):} Widely used in the literature for its simplicity and because it is optimal under the isotropic channel model from Sec.~\ref{ssec:iso}~\cite{Chowdhury2016}.
                It is constructed with $\mathbf{A}_{\mathrm{ED},k}\triangleq\id{N}/\trace(\mathbf{\Lambda}_k)$.
                \item \textit{High \acrshort{snr} (\acrshort{hsnr}):} High \acrshort{snr} approximation of the \acrshort{ml} detector for general fading corresponding to the case studied in Sec.~\ref{ssec:general}.
                Its associated weighting matrix is $\mathbf{A}_{\mathrm{HSNR},k}\triangleq\mathbf{\Lambda}_k^{-1}/N$.
                \item \textit{Assisted best quadratic unbiased estimator (\acrshort{abque}):}
                Decision-directed implementation of the \textit{\acrfull{bque}} (see~\cite{VilaInsa2024}).
                The output $\{\widehat{x}_k\}$ of a first hard decision from \acrshort{ed} is used to compute the weighting matrices of \acrshort{abque}, defined as
                $\mathbf{A}_{\mathrm{ABQUE},k}\triangleq\mathbf{\Lambda}_k\covmat{\bsf{r}\vert \widehat{x},k}^{-2}\bigm/\|\mathbf{\Lambda}_k\covmat{\bsf{r}\vert \widehat{x},k}^{-1}\|_{\frob}^2$.
                Then, a second detection round is performed on the received data with them.
            \end{enumerate}
            The complexity of detecting a codeword with any of these schemes (without the pre-processing) is $O(KN)+O(K\ln K)$, which does not scale with $\vert\mathcal{X}\vert$, contrary to \acrshort{ml} detection.
            This is especially relevant in the context of \acrshort{pm}, whose alphabet size can grow as fast as factorially.
            A summary of detection complexity is given in Table~\ref{tab:detectors_comparison}.
            Although the \acrshort{ed} is less complex, its performance is several orders of magnitude worse than the other detectors, up to the point of being unusable in practice.
            Moreover, the \acrshort{hsnr} and \acrshort{abque} detectors present the same complexity, but the latter's performance is superior, as it is shown next.

            \begin{table}
                \renewcommand{\arraystretch}{1.2}
                \centering
                \caption{Summary of detection schemes complexity}
                \vspace*{-1ex}
                \label{tab:detectors_comparison}
                \begin{tabular}{c|c}
                    Detector & Complexity\\ \hline\hline
                    \acrshort{ml} & $O(KN^2)+O(\lvert\mathcal{X}\rvert KN)$\\ \hline
                    \acrshort{ed} & $O(KN)+O(K\ln K)$\\ \hline
                    \acrshort{hsnr} \& \acrshort{abque} & $O(KN^2)+O(KN)+O(K\ln K)$ 
                \end{tabular}
                \vspace*{-1.25em}
            \end{table}
            
    \section{Numerical results}
        \label{sec:numerical_results}
        In this section the performance of the proposed low-complexity schemes against \acrshort{ml} detection is assessed.
        A \acrshort{pim} with $K=32$ and $L=4$ under uniform policy is employed: $K_l=K/L=8$, unless stated otherwise.\footnote{These values are limited by the maximum \acrshort{ml} complexity that can be simulated in a reasonably sized computation server. For instance, the number of trellis states required for the Viterbi algorithm for $K=64$ surpasses $8\cdot10^5$.}
        This configuration yields a \acrshort{se} of $\mathrm{R}_{32}(\{8\})\approx1.765\,\mathrm{bpcu}$, which amounts to a code rate of 0.8825 (the uncoded \acrshort{se} is $2\,\mathrm{bpcu}$).
        Symbols are drawn from a uniform unipolar 4-\acrshort{ask} constellation with unit average power.
        While a tailored constellation and nonuniform partition are expected to yield better performance~\cite{Han2022, marti_2024_constellation, AbouFaycal2001}, using a standard constellation and uniform partition allows to focus the discussion on the detector design, and is a reasonable choice in the context of low-complexity transmitters.
        Fading is assumed identically distributed across frequency bands (\ie $\covmat{\bsf{h},k}\equiv\covmat{\bsf{h}}$) under \textit{exponential correlation}~\cite{VilaInsa2024}: $[\covmat{\bsf{h}}]_{m,n}=\rho^{\lvert n-m\rvert}$, being $0\leq\rho\leq1$ the correlation factor.
        The \acrshort{ml} detector used as a benchmark is implemented with the Viterbi algorithm on a trellis with a total of 6561 states.
        Since bit coding is outside the scope of this paper, the metric considered herein is the average \acrfull{ser}, \ie the probability of incorrectly detecting each $x_k$.

        \begin{figure}[t]
            \centering
            \includegraphics[width=\linewidth]{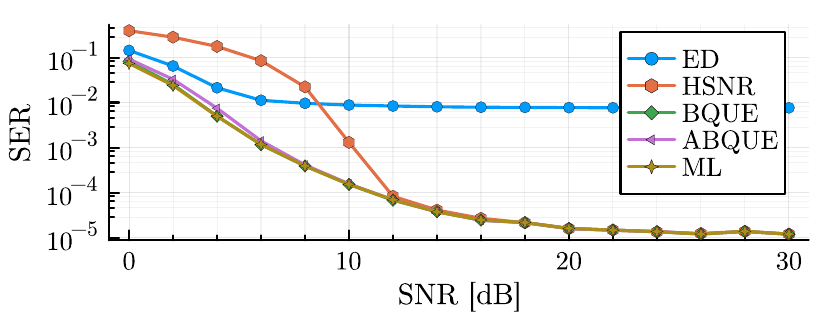}
            \vspace{-2em}
            \caption{\acrshort{ser} in terms of \acrshort{snr} for $N=64$ and $\rho=0.7$.}
            \label{fig:snr}
            \vspace{1em}
            \includegraphics[width=\linewidth]{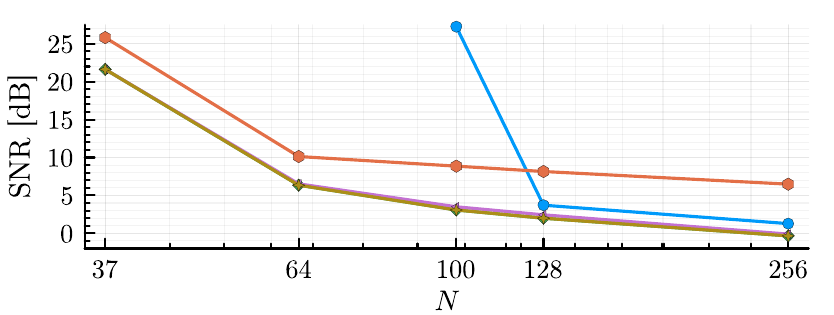}
            \vspace{-2em}
            \caption{Required \acrshort{snr} to achieve $\mathrm{SER}=10^{-3}$ in terms of $N$, for $\rho=0.7$.}
            \label{fig:ant}
            \vspace{1em}
            \includegraphics[width=\linewidth]{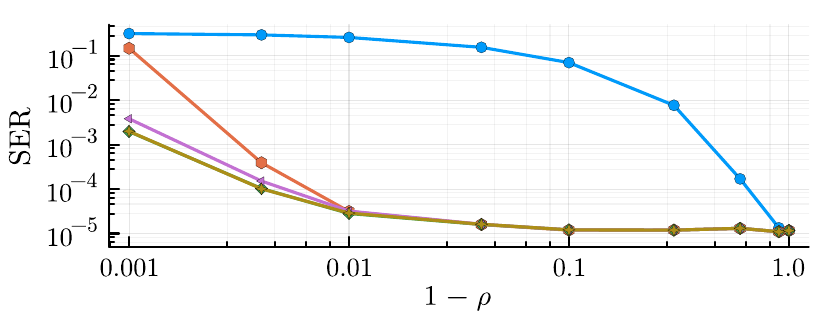}
            \vspace{-2em}
            \caption{\acrshort{ser} in terms of fading correlation for $N=64$ and $\snr=\SI{30}{\decibel}$.}
            \label{fig:corr}
            \vspace{1em}
            \includegraphics[width=\linewidth]{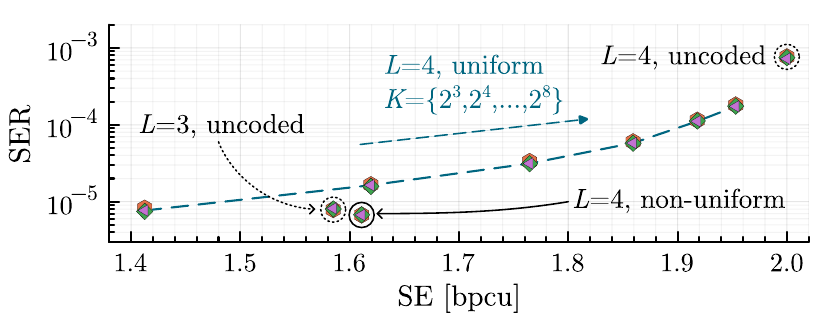}
            \vspace{-2em}
            \caption{\acrshort{ser} for various $K$ and partitioning policies with $N=64$, $\snr=\SI{15}{\decibel}$ and $\rho=0.7$.}
            \label{fig:rate}
            \vspace{-1.5em}
        \end{figure}

        In Fig.~\ref{fig:snr}, the performance of all detectors is displayed in terms of \acrshort{snr}.
        A first examination reveals a fundamental error floor at high \acrshort{snr} for all four detectors.
        It is expected in noncoherent systems for which different codewords induce the same signal subspace at the receiver~\cite{VilaInsa2025}.
        Analogous phenomena have been observed in other energy-based noncoherent schemes~\cite{VilaInsa2024,Jing2016}.
        \acrshort{ed} shows a higher error floor than the rest due to it being used under non-isotropic fading and not fully exploiting the channel correlation structures.
        \acrshort{hsnr} suffers a severe performance degradation at low and moderate \acrshort{snr} (\textless \SI{12}{\decibel}), but reaches \acrshort{ml} levels of \acrshort{ser} otherwise, when the approximation upon which it is built becomes valid.
        The most noteworthy aspect of Fig.~\ref{fig:snr} is that \acrshort{abque} displays near-\acrshort{ml} performance across all \acrshort{snr} levels in spite of its reduced complexity.
        It only suffers a slight penalty at very low \acrshort{snr} due to the poor performance of its first \acrshort{ed} round.

        Fig.~\ref{fig:ant} represents the minimum \acrshort{snr} required by each detector to achieve $\mathrm{SER}=10^{-3}$, for various numbers of antennas.
        The existence of the error floor previously discussed (which lowers for higher $N$ values~\cite{VilaInsa2024}) prevents detectors from reaching the objective \acrshort{ser} when $N$ is below a certain threshold.
        Unsurprisingly, \acrshort{ed} demands the largest $N$ of the four.
        \acrshort{abque} reaches \acrshort{ml} performance for $N\leq64$, but requires a slightly higher \acrshort{snr} when using more antennas.
        Similar performance gaps have been observed in~\cite{VilaInsa2024}.
        Finally, \acrshort{hsnr} involves the highest \acrshort{snr} levels for large values of $N$.

        Fig.~\ref{fig:corr} depicts the impact of antenna correlation onto each detector.
        As expected, \acrshort{ed} suffers a severe performance penalty with a small increase in $\rho$, becoming inoperative as soon as $\rho>0.7$.
        The other three detectors perform similarly for $\rho\leq0.99$, with graceful degradation through this wide range of correlation regimes.
        When $\rho>0.99$, the \acrshort{ser} of \acrshort{hsnr} starts to rapidly deteriorate.
        The performance loss for \acrshort{ml} and \acrshort{abque} is slower and their associated \acrshort{ser} values only diverge noticeably at extreme correlation values.
        Once again, a poor first \acrshort{ed} round negatively impacts \acrshort{abque}.

        Finally, in Fig.~\ref{fig:rate}, we illustrate the trade-off between \acrshort{se} and robustness of the proposed \acrshort{pim} scheme in terms of $K$ and partition policy.
        The \acrshort{ser} corresponding to the \acrshort{ed} is not provided since it is consistently outperformed by the other detectors.
        The one corresponding to the \acrshort{ml} detector is omitted as well due to computational limitations, and instead the performance of the genie-aided \acrshort{bque} detector is used as its benchmark~\cite{VilaInsa2024}.
        As stated in Sec.~\ref{ssec:pim}, \acrshort{se} increases with the number of subcarriers, at the cost of increased \acrshort{ser}.
        Such trend is displayed with a dashed teal line for $L=4$ and uniform policy.
        Ultimately, for $K\to\infty$, this modulation approaches the performance of the uncoded scheme, in which there is no \acrshort{se} penalty nor coding gain.

        In order to illustrate the potential advantages of using a nonuniform policy, in Fig.~\ref{fig:rate} we provide the \acrshort{ser}/\acrshort{se} of an \textit{ad hoc} code with $L=4$ and $K=30$, constructed from partition sizes $\{K_l\}=\{12,9,6,3\}$.
        This simple code outperforms the uncoded scheme for $L=3$ (both in terms of \acrshort{ser} and of \acrshort{se}), and also the uniform partition schemes for $L=4$, confirming the benefits of using the \acrshort{pim} scheme as a way of obtaining a coding gain~\cite{Fazeli2022}.
        Furthermore, if \acrshort{csi} is available at the transmitter, significant gains can be achieved by taking it into account in the constellation and partition design processes~\cite{marti_2024_constellation}.
        
    \section{Conclusion}
        In this work, a near-\acrshort{ml} low-complexity detector for noncoherent \acrshort{ofdm}-\acrshort{pim} has been presented.
        By leveraging the channel hardening property of a massive \acrshort{simo} scheme, we have derived a one-shot receiver that exploits statistical \acrshort{csi} to accurately compute second order statistics from data, which are then sorted to obtain the codeword decision.
        Numerical simulations have validated the error performance of this approach, reaching \acrshort{ml} levels in a variety of fading regimes with a fraction of its computational cost.

        Different lines of research may stem from ideas presented in this work.
        Of particular interest is the optimization of partition sizes and energy levels to provide enhanced \acrshort{se}/diversity trade-offs~\cite{Fazeli2022}.
        A related topic is the analysis of design criteria to trim the alphabet size to further improve its robustness to detection errors.
        On a final note, the \acrshort{pim} structure presented herein is specially suitable for hierarchical and multi-resolution~\cite{Seddik2017} systems.
        Combined coherent-noncoherent schemes based on \acrshort{pim} should be explored further.
        
    \appendix
        In order to prove $\lim_{K\to\infty} \mathrm{R}_{K}(\{K_l\}) = \mathrm{H}(\{p_l\})^-$ (\ie from below) we leverage the following bounds on the log-factorial function derived from the \textit{Stirling's series}~\cite[Ch.~27]{spivak_calculus_1994}:
        \begin{equation}
           \overbracket[1pt]{\log\tfrac{\sqrt{2\pi\alpha}\alpha^{\alpha}}{\euler^{\alpha}}+\tfrac{\log\euler}{12\alpha+1}}^{g_{\text{low}}(\alpha)} < \log(\alpha!) < \overbracket[1pt]{\log\tfrac{\sqrt{2\pi\alpha}\alpha^{\alpha}}{\euler^{\alpha}}+\tfrac{\log\euler}{12\alpha}}^{g_{\text{up}}(\alpha)}\label{eq:bounds}
        \end{equation}
        which are asymptotically tight as $\alpha\to\infty$.
        Moreover, the terms $\tfrac{\log\euler}{12\alpha+1}$ and $\tfrac{\log\euler}{12\alpha}$ vanish asymptotically and the bounds coincide.
        Thus, an upper bound for the \acrshort{se} in~\eqref{eq:rate} is:
        \begin{equation}\label{eq:rate_bounds}
            \begin{aligned}
                \textstyle\mathrm{R}_{K}(\{K_l\}) < \frac{1}{K} \bigl(g_{\text{up}}(K) - \sum_{l=1}^L g_{\text{low}}(K_l)\bigr).
            \end{aligned}
        \end{equation}
        Substituting~\eqref{eq:bounds} onto it yields
        \begin{equation}
            \begin{aligned}
                &\textstyle\mathrm{R}_{K}(\{K_l\})< \mathrm{H}(\{p_l\}) \\
                &\textstyle\quad+\frac{1}{K}\bigl(\frac{\log(2\pi K)}{2} + \frac{\log\euler}{12K} - \sum_{l=1}^{L} \frac{\log(2\pi K_l)}{2}+\frac{\log\euler}{12K_l+1}\bigr)
            \end{aligned}
        \end{equation}
        which tends to $\mathrm{H}(\{p_l\})$ for $K\to\infty$.
        In addition, $\exists M>0$ such that $\forall K > M$,
        \begin{equation}
            \textstyle\frac{\log(2\pi K)}{2} + \frac{\log\euler}{12K} - \sum_{l=1}^{L} \frac{\log(2\pi K_l)}{2}+\frac{\log\euler}{12K_l+1} \leq 0,
        \end{equation}
        since $\log(x)$ is subadditive for $x\geq1$.
        Hence, $\mathrm{R}_{K}(\{K_l\}) \leq \mathrm{H}(\{p_l\})$ and $\lim_{K\to\infty} \mathrm{R}_{K}(\{K_l\}) = \mathrm{H}(\{p_l\})^-$.

        To maximize the \acrshort{se} (defined in~\eqref{eq:rate}) in terms of $\{K_l\}$ in the non-asymptotic case, the following optimization problem must be solved with Lagrange multipliers:
        \begin{equation}
            \textstyle \argmin_{\{K_l\}} \sum_{l=1}^L \log\Gamma(K_l+1)\,\text{ s.t.}\,\sum_{l=1}^LK_l=K, \label{eq:opt}
        \end{equation}
        where $\Gamma(\boldsymbol{\cdot})$ is the gamma function~\cite[Ch.~19]{spivak_calculus_1994}.
        By the \textit{Bohr-Mollerup theorem}~\cite[Ch.~19]{spivak_calculus_1994}, it is known that $\Gamma(x)$ is log-convex for $x>0$.
        Then, by~\cite[Proposition~3.C.1]{Marshall2011}, the sum of convex functions~\eqref{eq:opt} is minimized by the uniform policy: $K_l=K/L$ for all $l$.
        The resulting \acrshort{se} of such alphabet is $\mathrm{R}_{K}(L) = \frac{1}{K} (\log(K!) - L\log(\frac{K}{L}!))$.
    
    \bibliography{IEEEabrv, refs}
\end{document}